\documentclass[api,pof,11pt]{revtex4-1}

\usepackage{graphicx}
\usepackage{dcolumn}
\usepackage{bm}
\usepackage{natbib}                                                             
\bibpunct{(}{)}{,}{s}{,}{,}

\begin{document}

\preprint{APS/123-QED}

\title{Exact coherent structures in a\\ reduced model of parallel shear flow}

\author{C\'edric Beaume}
 \email{ced.beaume@gmail.com}
\affiliation{Department of Physics, University of California, Berkeley CA 94720, USA}

\author{Gregory P. Chini}
 \email{greg.chini@unh.edu}
\affiliation{Department of Mechanical Engineering \& Program in Integrated Applied Mathematics, University of New Hampshire, Durham NH 03824} 
  
\author{Edgar Knobloch}
 \email{knobloch@berkeley.edu}
\affiliation{Department of Physics, University of California, Berkeley CA 94720, USA}

\author{Keith Julien}
 \email{keith.julien@colorado.edu}
\affiliation{Department of Applied Mathematics, University of Colorado at Boulder, Boulder CO 80309}

\date{\today}

\begin{abstract}
A reduced description of shear flows consistent with the Reynolds number scaling of lower-branch exact coherent states in plane Couette flow [J. Wang et al., Phys. Rev. Lett. 98, 204501 (2007)] is constructed. Exact time-independent nonlinear solutions of the reduced equations corresponding to both lower and upper branch states are found for Waleffe flow [F. Waleffe, Phys. Fluids 9, 883--900 (1997)].  The lower branch solution is characterized by fluctuations that vary slowly along the critical layer while the upper branch solutions display a bimodal structure and are more strongly focused on the critical layer.  The reduced model provides a rational framework for investigations of subcritical spatiotemporal patterns in parallel shear flows.

\end{abstract}

\pacs{Valid PACS appear here}

\maketitle

Exact nonlinear solutions of the equations describing the evolution of simple parallel shear flows have proved to be of immense value.\citep{Kawahara12} 
The existence of these solutions exposes the basic mechanism underlying self-sustained structures in shear flows and ultimately may shed light on the 
occurrence of subcritical turbulence in such flows. Despite notable successes\citep{Nagata90,Clever97,Waleffe97,Gibson08} the computation of such 
``exact coherent states/structures" (ECS) remains difficult because they are three-dimensional (3D) and disconnected from the 
structureless base shear flow.  
In this paper we propose a formulation that overcomes these difficulties.  This approach differs in certain important aspects from the pioneering analysis 
of Hall \& Sherwin\citep{Hall10} and builds on earlier work by the authors.\citep{CJK09,BeaumeGFD12} Specifically, we derive a simplified 
version of the governing partial differential equations (PDEs) that, as in the work of Hall \& Sherwin, yields an asymptotically exact description of lower 
branch states in the limit $Re\rightarrow\infty$, where $Re$ is a suitably defined Reynolds number.  Unlike these authors, we propose a \emph{composite} 
multiscale PDE model that is uniformly valid over the entire spatial domain.  Our derivation highlights the underlying PDE structure associated with the 
formation of ECS and, although not pursued here, also reveals how slow streamwise modulation of the mean (streamwise-averaged) and fluctuation 
(streamwise-varying) fields may be consistently incorporated. We solve the resulting equations by an iterative scheme, each step of which requires the 
solution of a two-dimensional problem only.  While Hall \& Sherwin\citep{Hall10} and, more recently, Blackburn \emph{et al.}\citep{Blackburn13} focus 
on plane Couette flow, we demonstrate the method on a simple sinusoidal, body-forced shear flow with stress-free boundaries that we call Waleffe flow.\citep{Waleffe97}  
The detailed structure of the ECS we compute necessarily differs from that of ECS in Couette flow.  Remarkably, we show that for Waleffe flow the method not only 
captures the lower branch states for which it was developed, but also \emph{upper branch} states:  in spite of the large $Re$ formulation, the asymptotics prove 
sufficiently robust to capture the saddle-node bifurcation giving rise to these solutions.  For the given domain size, this bifurcation occurs at $Re\approx 136$ and we are able to numerically continue both branches from this value to $Re>2000$. 


We consider incompressible flow driven by a streamwise body force that varies sinusoidally in the wall-normal ($y$) direction,\citep{Waleffe97}
\begin{eqnarray}
\label{wf1}&\partial_t {\bf u} + \left( {\bf u} \cdot \nabla \right) {\bf u} = - \nabla p + \frac{1}{Re} \nabla^2 {\bf u} + \frac{\sqrt{2} \pi^2}{4 Re} \sin (\frac{1}{2}\pi y){\bf {\hat x}},\\
\label{wf2}&\nabla \cdot {\bf u} = 0,
\end{eqnarray}
subject to stress-free boundary conditions at stationary walls located at $y=\pm1$,
\begin{equation} 
\label{boundary}
\partial_yu=v=\partial_yw=0.
\end{equation}
Here $Re\equiv UL_y/\nu$ is the Reynolds number, where $L_y$ is the channel half-width and $U$ is the root-mean-square velocity of the base flow given in dimensionless form by $(u,v,w)=(\sqrt{2}\sin(\pi y/2),0,0)$, hereafter referred to as Waleffe flow. Like the better studied plane Couette flow, Waleffe flow is linearly stable for all $Re$ but may be unstable to finite amplitude perturbations. The codimension-one states on the boundary separating the basin of attraction of Waleffe flow from that of the upper branch states are called edge states\cite{Skufca06} and are found on the lower branch. These nonlinear states are maintained against decay by the self-sustaining instability mechanism elucidated by Waleffe.\cite{Waleffe97}

Given the occurrence of streamwise streaks and rolls that typify ECS in shear flows, we decompose the velocity vector into a streamwise component and a perpendicular vector, i.e., $\mathbf{v}=(u,\mathbf{v}_\bot)$, where $\mathbf{v}_\bot=(v,w)$, and posit appropriate asymptotic expansions for the various fields.  To this end, we are motivated in part by the scaling behavior identified by Wang \emph{et al.}\citep{Wang07} for lower-branch ECS in Couette flow.  As in the work of Hall \& Sherwin\citep{Hall10} and as first demonstrated by Waleffe,\citep{Waleffe97} the rolls comprising the streamwise-averaged flow in the perpendicular plane are weak, $\mathit{O}(\epsilon)$ where $\epsilon\equiv 1/Re$, relative to the deviation of the streamwise-averaged streamwise flow from the base laminar profile (i.e., relative to the streaks). A \emph{closed} and asymptotically consistent reduced model may be obtained by further positing that the (streamwise-varying) fluctuations are similarly weak relative to the mean streamwise flow, an assumption consistent with the scaling behavior reported by Wang \emph{et al.}\citep{Wang07} We suppose that all fields are functions of $(x,X,y,z,t,T)$, where $X\equiv\epsilon x$ and $T\equiv \epsilon t$ are slow scales,\citep{CJK09} and write
\begin{eqnarray}
u\sim\bar{u}_0+\epsilon\left(\bar{u}_1+u_1'\right)+\ldots,\;
\mathbf{v}_\bot\sim\epsilon\left(\bar{\mathbf{v}}_{1\bot}+\mathbf{v}_{1\bot}'\right)+\ldots,\;
p\sim\bar{p}_0+\epsilon\left(\bar{p}_1+p_1'\right)+\epsilon^2\left(\bar{p}_2+p_2'\right)+\ldots,\quad{}
\end{eqnarray}
where an overbar denotes a ``fast" $(x,t)$ average and a prime denotes a fluctuation with zero fast mean.  Substituting these expansions into the multiscale versions of Eqs.~(\ref{wf1})--(\ref{wf2}), collecting terms at like order in $\epsilon$, and parsing the resulting equations into mean and fluctuating components yields the following asymptotically-reduced, multiscale PDE system:
\begin{eqnarray}
\partial_T\bar{u}_0 + \bar{u}_0\partial_X\bar{u}_0
+ \left(\bar{\mathbf{v}}_{1\bot}\cdot\nabla_\bot\right)\bar{u}_0
&=&-\partial_X\bar{p}_0\,+\,\frac{\sqrt{2}\pi^2}{4}\sin\left(\frac{\pi y}{2}\right)+\nabla_\bot^2\bar{u}_0,\label{U0barEQN}\\
\partial_T\bar{\mathbf{v}}_{1\bot}+\partial_X\left[\bar{u}_0\bar{\mathbf{v}}_{1\bot}\right]
+\nabla_\bot\cdot\left[\bar{\mathbf{v}}_{1\bot}\bar{\mathbf{v}}_{1\bot}+
\overline{\mathbf{v}_{1\bot}'\mathbf{v}_{1\bot}'}\right]&=&-\nabla_\bot\bar{p}_2
+ \nabla_\bot^2\bar{\mathbf{v}}_{1\bot},\label{Vp1barEQN}\\
\partial_X\bar{u}_0\,+\,\nabla_\bot\cdot\bar{\mathbf{v}}_{1\bot}&=&0,\label{CONTbarEQN}
\end{eqnarray}
which govern the mean dynamics, and  
\begin{eqnarray}
\partial_t u_1' + \bar{u}_0\partial_x u_1' 
+ \left(\bar{\mathbf{v}}_{1\bot}'\cdot\nabla_\bot\right)\bar{u}_0
&=&-\partial_x p_1'\,+\,\epsilon\nabla_\bot^2u_1',\label{U1primeEQN}\\
\partial_t \mathbf{v}_{1\bot}' + \bar{u}_0\partial_x\mathbf{v}_{1\bot}'
&=&-\nabla_\bot p_1'\,+\,\epsilon\nabla_\bot^2\mathbf{v}_{1\bot}',\label{Vp1primeEQN}\\
\partial_x u_1'\,+\,\nabla_\bot\cdot\mathbf{v}_{1\bot}'&=&0,\label{CONTprimeEQN}
\end{eqnarray}
which govern the fluctuating fields.  Here, $\nabla_\bot$ is the gradient operator in the $y$--$z$ plane. Note that $\bar{p}_0=\bar{p}_0(X,T)$ is set to zero for Waleffe and Couette flow, but may be retained for flows driven by externally-imposed pressure gradients, such as plane Poiseuille flow. We emphasize that Eqs.~(\ref{U0barEQN})--(\ref{CONTprimeEQN}) comprise a \emph{closed} reduced system; the usual closure issues resulting from averaging do not arise here owing to our ability to exploit scale separation.  Physically, the averaged equations constrain the slow temporal and streamwise evolution of the streaks ($\bar{u}_0$) and rolls ($\bar{\mathbf{v}}_{1\bot}$). The presence of an effective Reynolds number equal to unity and the elimination of fast streamwise and temporal variation in these equations facilitates both time-stepping and the computation of equilibrium ECS in comparison with Eqs.~(\ref{wf1})--(\ref{wf2}) at $Re\gg1$.  Further savings accrue if the slow streamwise ($X$) variation is suppressed, as in our computations here, since the averaged equations are then spatially 2D.


Presuming fluctuation gradients remain $\mathit{O}(1)$, the fluctuating fields themselves evolve in accord with the equations governing the \emph{inviscid secondary stability} of streamwise streaks. 
As explicitly demonstrated by Hall \& Sherwin,\citep{Hall10} the fluctuation fields, which are necessarily steady (i.e., neutrally stable) for equilibrium ECS, exhibit a critical layer structure along the isosurface $\bar{u}_0(y,z)$=$0$.  In the neighborhood of the critical layer, fluctuation gradients are large, resulting in a distinct leading-order dominant balance of terms involving diffusion.  However, we choose to avoid the intricacies associated with carrying out a systematic matched asymptotic analysis to address the critical layer singularity.\citep{Hall10}  Instead we retain the formally small perpendicular diffusion terms in Eqs.~(\ref{U1primeEQN})--(\ref{Vp1primeEQN}), which are then uniformly valid over the entire spatial domain. Retention of these terms may be justified by appeal to the method of \emph{composite asymptotic equations}, as in Ref.\cite{GiannettiLuchini06}

It is important to note that the fluctuation equations do not mix $x$ modes, a fact that we exploit in our computations of ECS for Waleffe flow using the reduced system.  In fact, in accord with the scalings found by Wang \emph{et al.},\citep{Wang07} we retain only a \emph{single} streamwise Fourier mode for each fluctuation field,
$[u_1',\mathbf{v}_{1\bot}',p_1'](x,y,z,t)=[u_1,\mathbf{v}_{1\bot},p_1](y,z,t)e^{i\alpha x}+c.c.$,
where $c.c.$ denotes complex conjugate and $\alpha=2\pi/L_x$ is the fundamental streamwise wavenumber.
Before describing the computation of \emph{streamwise uniform} ECS, we remark that in long domains a nearly continuous band of modes with similar streamwise wavenumbers will be neutral or very weakly damped.  Hence, a linear superposition of these fluctuation modes will naturally induce a \emph{slowly-varying} envelope, $A(X,T)$ say, that will in turn drive slow streamwise modulations of the mean fields through the Reynolds stress divergence term in Eq.~(\ref{Vp1barEQN}). If realized, this multiscale coupling may provide a mechanism for streamwise localization of ECS in a variety of plane parallel shear flows, further attesting to the value of the reduced PDE structure identified here.

With $X$ derivatives suppressed, Eqs.~(\ref{U0barEQN})--(\ref{CONTbarEQN}) can be further simplified by introducing a streamwise-invariant streamfunction $\phi_1(y,z)$ so that $\bar{v}_1 = -\partial_z \phi_1$ and $\bar{w}_1 = \partial_y \phi_1$, and the corresponding streamwise vorticity $\omega_1 = \nabla_{\perp}^2 \phi_1$, resulting in the following set of equations:
\begin{eqnarray}
\label{rdc1}&\partial_T u_0 + J(\phi_1,u_0) = \nabla_{\perp}^{2} u_0 + \frac{\sqrt{2} \pi^2}{4} \sin(\frac{1}{2}\pi y),\\
\label{rdc2}&\partial_T \omega_1 + J(\phi_1,\omega_1) + 2 (\partial_{yy}^2 - \partial_{zz}^2) \left( \mathcal{R}(v_1 w_1^*) \right) + 2 \partial_y \partial_z (w_1 w_1^* - v_1 v_1^*) = \nabla_{\perp}^2 \omega_1,
\end{eqnarray}
where $J(\phi_1,f) \equiv \partial_y \phi_1 \partial_z f - \partial_z \phi_1 \partial_y f$, $f^*$ denotes the complex conjugate of $f$, and $\mathcal{R}(f)$ denotes its real part: since $u_0'\equiv 0$ the overbar on the $\mathit{O}(1)$ streaky flow component has been omitted.  The fluctuation equations can be written 
in the more useful form
\begin{eqnarray}
\label{rdc3}&(\alpha^2 - \nabla_{\perp}^2) p_1 = 2 i \alpha (v_1 \partial_y u_0 + w_1 \partial_z u_0),\\
\label{rdc45}&\partial_t \mathbf{v}_{1\bot} + i \alpha u_0 \mathbf{v}_{1\bot} = -\nabla_\bot p_1 + \epsilon \nabla_{\perp}^2 \mathbf{v}_{1\bot}.
\end{eqnarray}
Note that $u_1$ is not required to close the equations although it may also be computed.  In the following these equations are solved subject to the stress-free boundary conditions
\begin{equation}
\partial_y u_0 = \omega_1 = \phi_1 = v_1 = \partial_y w_1 = 0, \quad {\rm at} \quad y=\pm 1.
\end{equation}

\begin{figure}
\centerline{\includegraphics[]{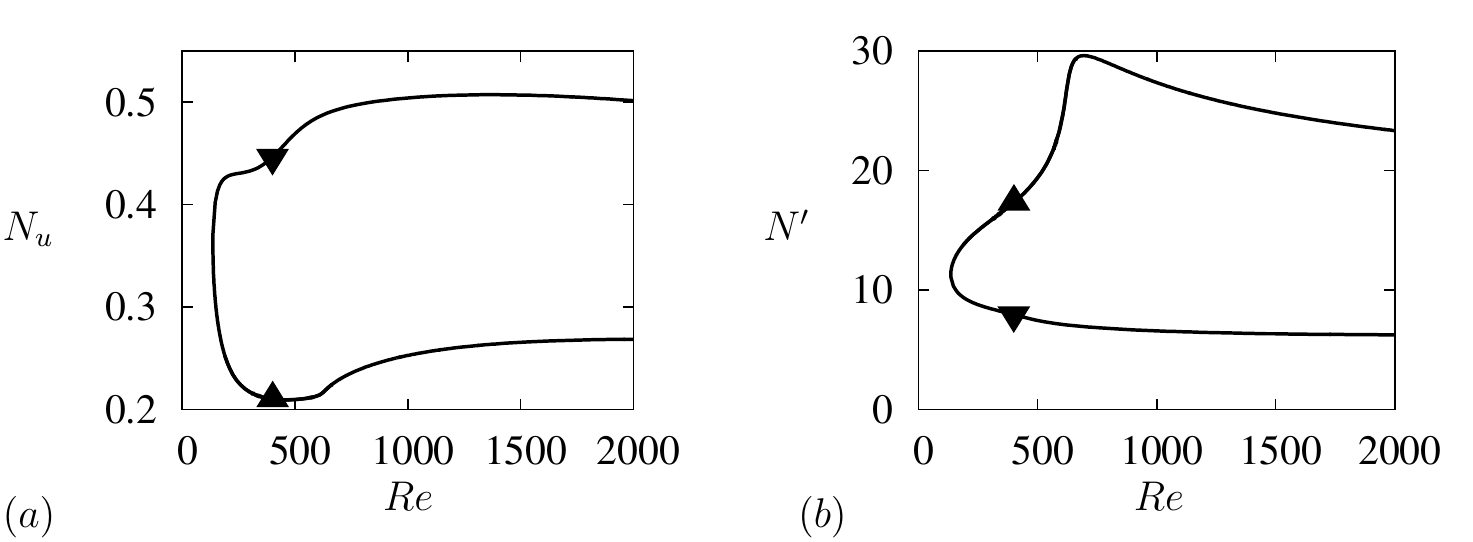}}
\caption{Bifurcation diagram showing the lower (downward triangle) and upper (upward triangle) branches of ECS as a function of the Reynolds number $Re$ in terms of (a) $N_u=\int_{\mathcal{D}} u_0^2 \, dy \, dz / \int_{\mathcal{D}} dy \, dz$,
(b) $N' = \int_{\mathcal{D}} (v_1^2+w_1^2) \, dy \, dz / \int_{\mathcal{D}} dy \, dz$. The branches are connected via a saddle-node bifurcation at $Re \approx 136$. Lower branch states are computed on a $32 \times 64$ mesh while upper branch states are computed on a $64 \times 128$ mesh. }
\label{bifdiag}
\end{figure}

Equations (\ref{rdc3})--(\ref{rdc45}) are homogeneous and quasilinear with solutions that depend on the slowly evolving streamwise velocity $u_0$. The solutions of these equations therefore either grow or decay. Since we are interested in stationary solutions of Eqs.~(\ref{wf1})--(\ref{wf2}) we use an iterative scheme consisting of two steps: searching for neutrally stable solutions of Eqs.~(\ref{rdc3})--(\ref{rdc45}) on the fast time scale $t$, and converging $u_0$ to a stationary state on the slow time scale $T$. We solve this problem on a two-dimensional domain $\mathcal{D}$ of size $2L_y \times L_z = 2 \times \pi$, where $L_z$ is an imposed period in the spanwise direction, and set $\alpha = 0.5$. In plane Couette flow this choice of domain leads to edge states with a single unstable direction.\citep{Schneider08} 
The computations are performed in spectral space using a mixed Fourier cosine/sine basis.
Once a steady nontrivial solution has been found numerical continuation in $Re$ is applied to trace out the whole solution branch.
For simplicity we impose the symmetry $[u,v,w](x,y,z) = [u,v,-w](x+L_x/2,y,-z)$ observed in the corresponding solutions in plane Couette flow,\citep{Schneider08} where $L_x= 4 \pi$ is the imposed period of the solution in the streamwise direction. The details of the iterative scheme used to solve this problem are nontrivial and will be described elsewhere\cite{tobesubmitted} together with details of the continuation scheme.


Figure \ref{bifdiag} shows the results in terms of $N_u \equiv \int_{\mathcal{D}} u_0^2 \, dy \, dz / \int_{\mathcal{D}} dy \, dz$, measuring the strength of the streaks, and $N'\equiv \int_{\mathcal{D}} (v_1^2+w_1^2) \, dy \, dz / \int_{\mathcal{D}} dy \, dz$, measuring the strength of the associated spanwise fluctuations $(v_1,w_1)$. These quantities are related to the kinetic energies per unit volume associated with these modes by $E_u = N_u/2$ and $E'=2 \pi N'/(\alpha Re^2)$. The figure shows that the reduced system captures not only the lower branch states for which it was developed but the upper branch states as well. The two branches connect via a saddle-node bifurcation at $Re \approx 136$.

\begin{figure}
\centerline{\includegraphics[]{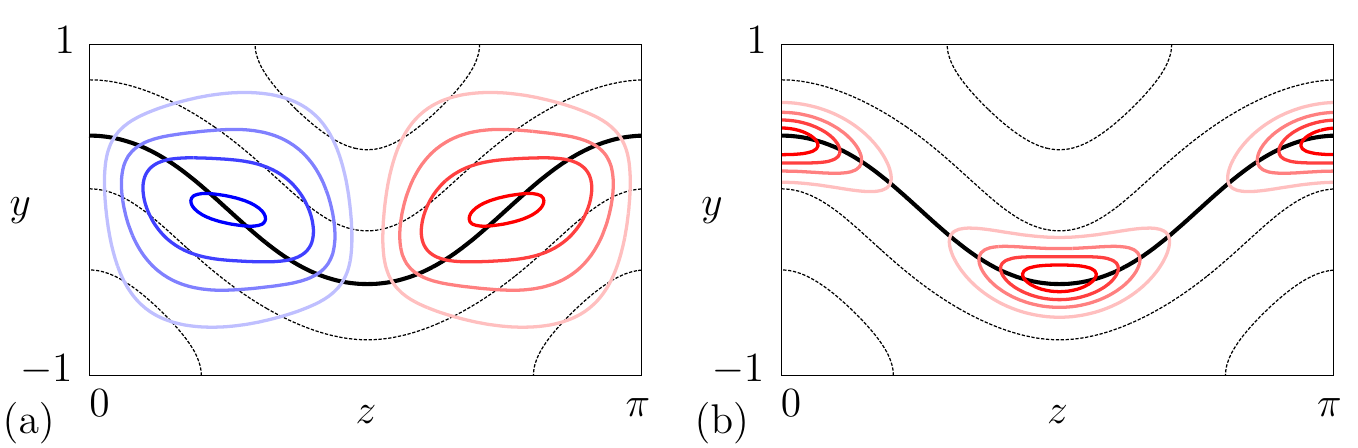}}
\caption{The lower branch solution at $Re \approx 1500$ represented by (a) contours of the streamwise streamfunction $\phi_1$ and (b) the quantity $||(v_1,w_1)||_{L_2}$, a measure of spanwise fluctuations. In each plot positive (negative) values are indicated in red (blue). The contour plots are superposed on the streak profile shown in black, with the thick solid line representing the critical layer $u_0 = 0$. All contours are equidistributed.}
\label{lower}
\end{figure}

\begin{figure}
\centerline{\includegraphics[]{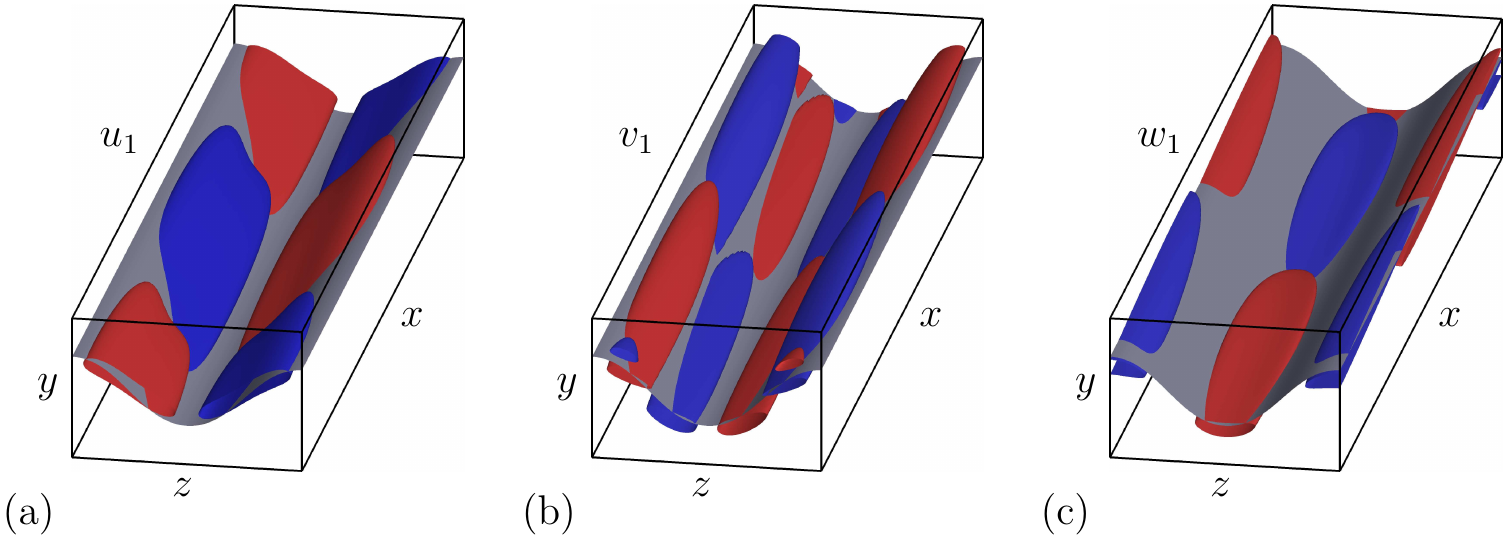}}
\caption{Three-dimensional rendition of the fluctuating flow on the lower branch solution at $Re \approx 1500$. The surfaces represented in color correspond to 
(a) $\frac{1}{2}\,{\rm max}|u_1|$, (b) $\frac{1}{2}\,{\rm max}|v_1|$, and (c) $\frac{1}{2}\,{\rm max}|w_1|$, with red (blue) representing positive (negative) values. The surface shown in grey represents the critical layer $u_0 = 0$.}
\label{3dlower}
\end{figure}

\begin{figure}
\centerline{\includegraphics[]{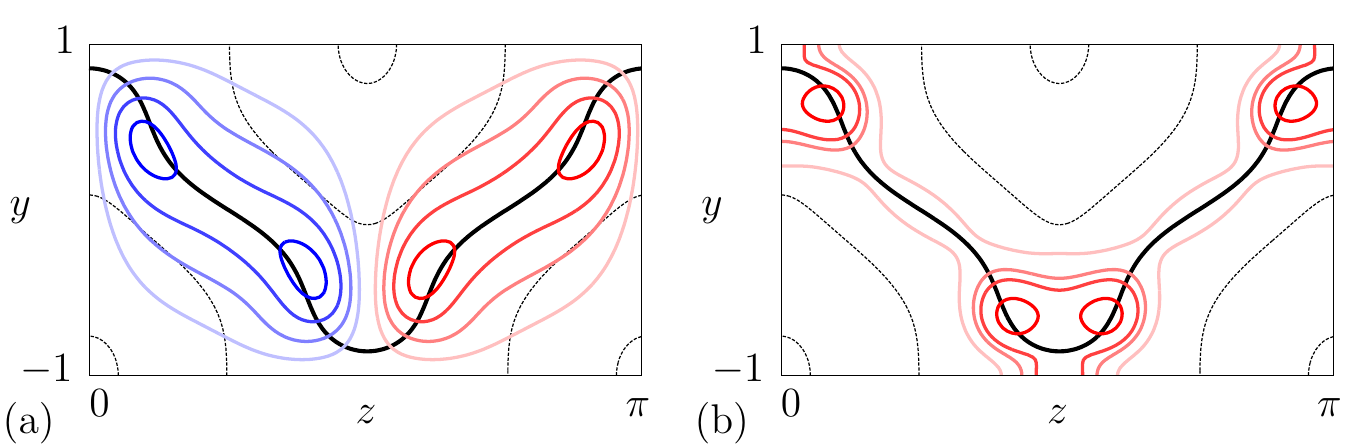}}
\caption{Same as Fig.~\ref{lower} but for the upper branch solution at $Re \approx 1500$.}
\label{upper}
\end{figure}

\begin{figure}
\centerline{\includegraphics[]{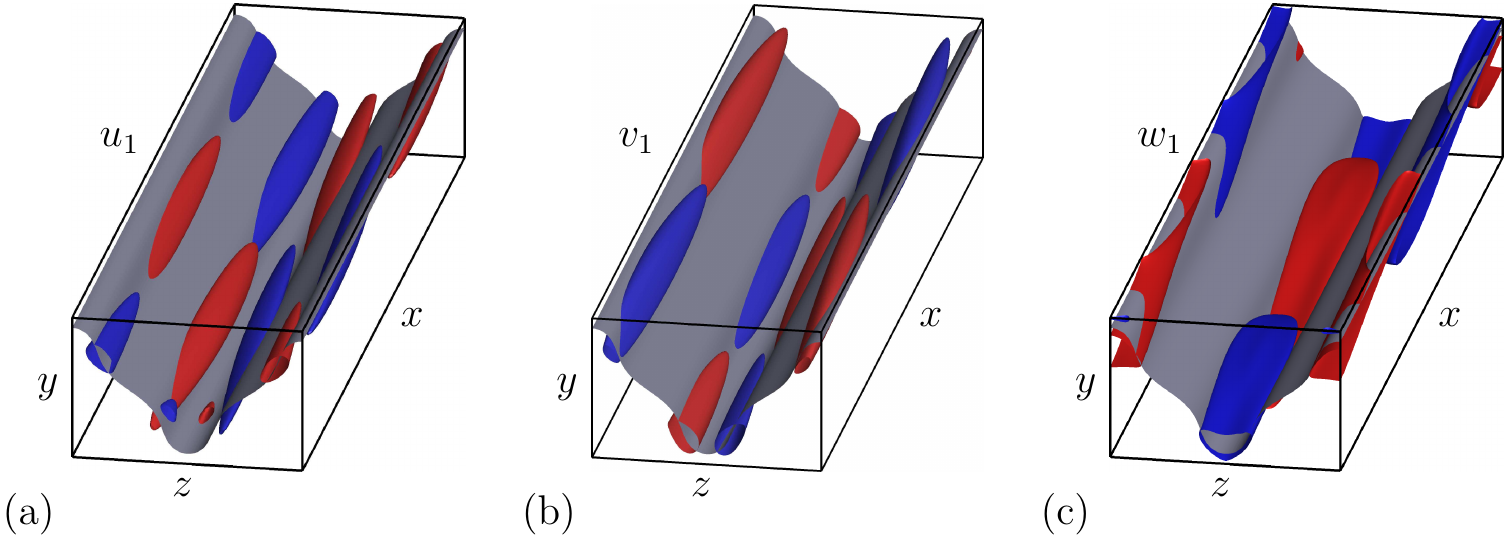}}
\caption{Same as Fig.~\ref{3dlower} but for the upper branch solution at $Re \approx 1500$. Intersections of the fluctuations with the walls at $y=\pm 1$ that can be observed in (c) are a consequence of the stress-free boundary conditions.}
\label{3dupper}
\end{figure}

Figure \ref{lower} shows streamwise-averaged representations of the lower branch solution at $Re \approx 1500$ while Fig.~\ref{3dlower} provides insight into the 3D structure of this solution. Figures \ref{upper} and \ref{3dupper} provide analogous representations of the upper branch solution at the same Reynolds number. The lower branch solution is characterized by a smoothly undulating critical layer that is maintained by two nearly circular rolls (Fig.~\ref{lower}(a)). This structure is supported by fluctuations that concentrate along a critical layer of $\mathit{O}(\alpha Re)^{-1/3}$ width.\cite{Maslowe86,Wang07,Hall10} Figure \ref{lower}(b) reveals that these fluctuations vary rapidly in the direction perpendicular to the critical layer with a much slower variation along it. The 3D representation in Fig.~\ref{3dlower} confirms these observations and sheds more light on the streamwise dynamics of the lower branch solution: the streamwise velocity fluctuation $u_1$ is concentrated in the regions of strong streamwise streamfunction $\phi_1$ (compare Fig.~\ref{lower}(a) with Fig.~\ref{3dlower}(a)) and therefore away from the extrema of the critical layer. In contrast, spanwise fluctuations $(v_1,w_1)$ accumulate at the extrema of the critical layer (Fig.~\ref{lower}(b)), a consequence of the incompressibility of the fluctuations (Eq.~(\ref{CONTprimeEQN})). At $x=0$ (defined arbitrarily as the front section in Fig.~\ref{3dlower}), the fluid in the region around the lowest (resp., highest) point of the critical layer tracks the critical layer from left to right (resp., right to left); the directions are reversed at locations displaced half a period in the streamwise direction.


In contrast with the nearly sinusoidal critical-layer profile exhibited by the lower branch solution, the critical layer associated with the upper branch solution is much more strongly 
deformed from the plane $y=0$, even approaching at its extrema the top and bottom walls.
This change of shape is a signature both of less coherent roll motions and of the splitting of each roll into a bimodal structure (Fig.~\ref{upper}(a)). This splitting moves the maximum of the streamwise streamfunction closer to the extrema of the critical layer to support its highly distorted profile. The resulting state resembles the upper branch solution in plane Couette flow found in Ref.\cite{Jimenez05} Figures \ref{3dupper}(a)--(c) show that the fluctuations associated with this state exhibit properties similar to those on the lower branch: the spanwise fluctuations $(v_1,w_1)$ are concentrated at the extrema of the critical layer with the streamwise velocity fluctuation $u_1$ expelled from these regions. However, the fluctuations also exhibit a bimodal structure with maximum values now located on either side of the critical layer extrema (Fig.~\ref{upper}(b)). This splitting serves to confine the critical layer in these regions, and leads to strong gradients in the fluctuation kinetic energy {\it along} the critical layer. 

We have described an asymptotic reduction procedure suggested by the lower branch scaling for plane Couette flow that has allowed us to compute both lower and upper branch solutions for Waleffe flow. This flow is easier to study because the boundary conditions are stress-free, enabling us to employ and refine a \emph{uniform} computational grid associated with a trigonometric basis in all coordinate directions.  A similar asymptotic approach has recently been used to obtain lower branch solutions to plane Couette flow\cite{Hall10,Blackburn13} but no upper branch states were reported, possibly because the upper branch states in these two flows scale differently with different boundary conditions.  In any case, the elimination of $\epsilon$ from the 
formulation of Hall and collaborators naturally precludes continuation in Reynolds number. 



Our lower branch solutions are qualitatively similar to those for plane Couette flow, but the upper branch solutions reveal properties heretofore unknown. These center on the appearance of the bimodal structure of both the streamwise rolls and the associated fluctuations. The accuracy of these solutions as ECS of the fully 3D problem is expected to improve with increasing Reynolds number. In future work we will report on the stability properties of these states and their relation to the ECS of the 3D problem. We conclude by reiterating the potential utility of the reduced PDEs:  because they capture the saddle-node bifurcation, they may provide a systematic yet simplified framework for spatiotemporal ECS pattern formation studies, and although they were derived for a parallel shear flow, the inclusion of slow streamwise variability suggests a systematic path for computing ECS in \emph{developing} flows, including boundary layers.

 
{\bf Acknowledgement}: This work was initiated during the 2008 NCAR Geophysical Turbulence Phenomena workshop in Boulder, CO, and was substantively continued during the Geophysical Fluid Dynamics Program at the Woods Hole Oceanographic Institution in 2012. 
The authors gratefully acknowledge support from the National Science Foundation under grant No.~DMS-1211953 (CB \& EK), No.~OCE-0934827 (GPC), and No.~OCE-0934737 (KJ).






\bibliography{wally}
\bibliographystyle{jasanum}

\end{document}